\documentstyle[12pt,worldsci]{article}
\begin{document}
\title{The $Q\bar{Q}$ Potential in the Color-Dielectric Formulation
of the Transverse Lattice}
\author{Bob Klindworth and Matthias Burkardt\\
{\em Department of Physics,
New Mexico State University, Las Cruces, NM 88003, U.S.A.}}
%\date{\today}
\maketitle
\setlength{\baselineskip}{2.6ex}

\vspace{0.7cm}
\begin{abstract}
The $Q\bar{Q}$ potential is calculated in the Color-Dielectric
Formulation of Transverse Lattice QCD. In such a formulation
an effective potential is used to enforce the $SU(N_c)$ symmetry
of the link fields. This effective potential is truncated at fourth
order. Surprisingly, there is striking rotational invariance in
the ground state $Q\bar{Q}$ potential as well as in
the first hybrid mode.
\end{abstract}
\vspace{0.7cm}
\section{Introduction}
Light front field theories have many advantages. First,
high energy experiments often involve large momentum transfers. Therefore,
it seems plausible that such experiments could be well-described in a
frame moving at the speed of light. Second, light front wavefunctions can
be regarded as correlation functions at equal light front time. From these
wavefunctions it is comparatively easy to extract important physical
observables like parton distribution functions, and these observables
have a direct physical interpretation in terms of light-front coordinates.

The transverse lattice is a marriage of light front field theory with the
very successful lattice gauge theory. One first defines the light front
variables $x^{\pm}=\frac{1}{\sqrt{2}}(x^0 \pm x^3)$ and keeps them
continuous. The remaining two directions ($x^1, x^2$) are discretized.
Like all light front theories, the transverse lattice lacks manifest
rotational invariance. For this reason calculating the $Q\bar{Q}$ potential
on the transverse lattice is a good way of testing the dynamics of our
model.

In previous work we calculated the $Q\bar Q$ potential using
2+1 dimensional Transverse Lattice QCD \cite{studs}.
In such a theory the lattice
gauge fields are matrices associated with the links between lattice
sites. Ideally, we would like to work with an effective potential that has
minima in $SU(N_c)$. However, as explained in Ref. [3], %\cite{dielectric} 
there are
difficulties with the Light Front formulation of such a theory.
Instead we use the Color-Dielectric formulation in which the link variables
are assumed to be smeared variables. This framework allows us to cover a
wider range of physical distances using relatively few degrees of freedom.

The results of the original 2+1 calculation exhibited surprising rotational
invariance. This was quite surprising because we used a very crude
effective potential, accepting only terms up to second order in the link
fields. In the current work we repeat the analysis from Ref. [2],
%reference \cite{studs},
including terms up to fourth order in the effective potential and
extending the calculation to 3+1 dimensions.

\section{The Effective Potential}
The purpose of the effective potential is to constrain the link fields
to be members of $SU(N_c)$. The terms in the
effective potential up to fourth order are \cite{dielectric}

\begin{eqnarray}
V_{eff}(U) & = & \mu^2 Tr  ( U_i U_i^\dagger ) +  
\frac{\lambda_1}{a N_c} Tr ( U_i U_i^\dagger U_i U_i^\dagger ) +
\frac{\lambda_2}{a N_c} Tr ( U_{i+1} U_i^\dagger U_{i+1} U_i^\dagger )
\nonumber \\
&   & +
\frac{\lambda_3}{a N_c^2} Tr ( U_i U_i^\dagger ) Tr ( U_i U_i^\dagger )
\label{eq:veff1}
\end{eqnarray}
\begin{figure}
\begin{centering}
\unitlength1.cm
\begin{picture}(5,5.5)(2,-7.7)
\includegraphics{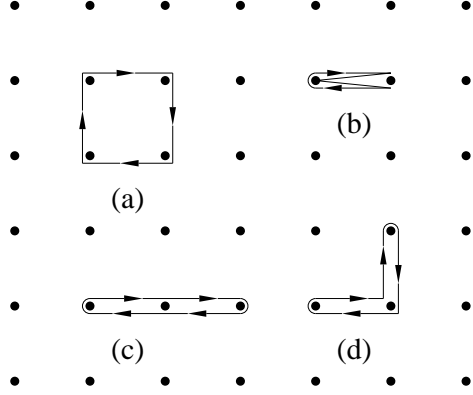}
\end{picture}
\caption{Relevant fourth order interactions in the effective potential for the
link fields. (a) canonical plaquette interaction, 
(b) local interaction with coupling $c_1$, (c) longitudinal
nonlocal interaction with coupling $c_L$, (d) transverse
nonlocal interaction with coupling $c_\perp$.
In $2+1$ dimensions only (b) and (c) contribute.
}
\label{fig:veff}
\end{centering}
\end{figure}

Our initial calculation truncated $V_{eff}$ at second order. 
The second and the fourth terms in this expansion do not contribute due
to our Fock Space truncation. We know that our truncation is valid in the
case of the ground state, so neglecting these terms seems plausible. It
is hoped that this is also valid for the lowest excited states.
Thus, the only fourth order term which contributes to our calculation is
the nonlocal interaction.

\begin{eqnarray}
V_{eff}(U) & = & \mu^2 Tr ( U_i U_i^\dagger ) + \nonumber 
\frac{\lambda_2}{a N_c} Tr ( U_{i+1} U_i^\dagger U_{i+1} U_i^\dagger )
\label{eq:veff2}
\end{eqnarray}

It was found in Ref. [3] %reference \cite{dielectric} 
that the
glueball spectrum could be fit to reliable Euclidean Lattice Monte
Carlo data as long as $\mu^2$ and $\lambda_2$ lie along a scaling
trajectory. In our calculation, we verified that points off of the
scaling trajectory give rise to potentials which are not rotationally
invariant.

\section{Results in 2+1 Dimensions}
With the effective potential fixed we calculated the $Q\bar{Q}$ Potential
as before using the 2+1 dimensional Transverse Lattice.
We used Discretized Light Cone Quantization \cite{pa:dlcq}. The main
differences when compared
to the calculation in Ref. [2] %\cite{studs} 
are the addition of a ``mass'' term
to the link fields and the inclusion of an attractive (Lorentz-) scalar
interaction between adjacent links. The resulting potential has an
anisotropy in the points with zero transverse separation resulting
from the fact that the non-local coupling does not act on states
whose transverse separation is less than two. Rotational invariance
is restored by introducing a new term in the effective potential.
\cite{dielectric2}
In momentum space this interaction has the form,

\begin{equation}
V_{contact}=%\frac{c_{contact}}{\sqrt{k^+k'^+}}
\frac{c_{contact}}{v^+} \frac{1}{\sqrt{k^+k'^+}}
\end{equation}

where $k^+$ and $k'^+$ are the incoming/outgoing momenta of the link
fields which interact with the external charges. Note that this interaction
acts only on the first and last link field in the 
``chain'' connecting the $Q\bar{Q}$ pair since it involves
the external charges. For the same reason, it does not conserve gluon
momentum. This interaction has a Lorentz structure similar to the
4-point interaction between fermions and bosons that arises when the
constrained component of the spinor field is eliminated. 

In addition to studying the ground state potential we can also study the
rotational invariance and qualitative shape of the first excited
state. Surprisingly, this state also exhibits striking rotational
invariance. This is surprising because higher Fock components, which
we omit, could contribute significantly to the excited states.
\begin{figure}
\begin{centering}
\unitlength1.cm
\begin{picture}(5,6)(-5.8,.8)
\includegraphics{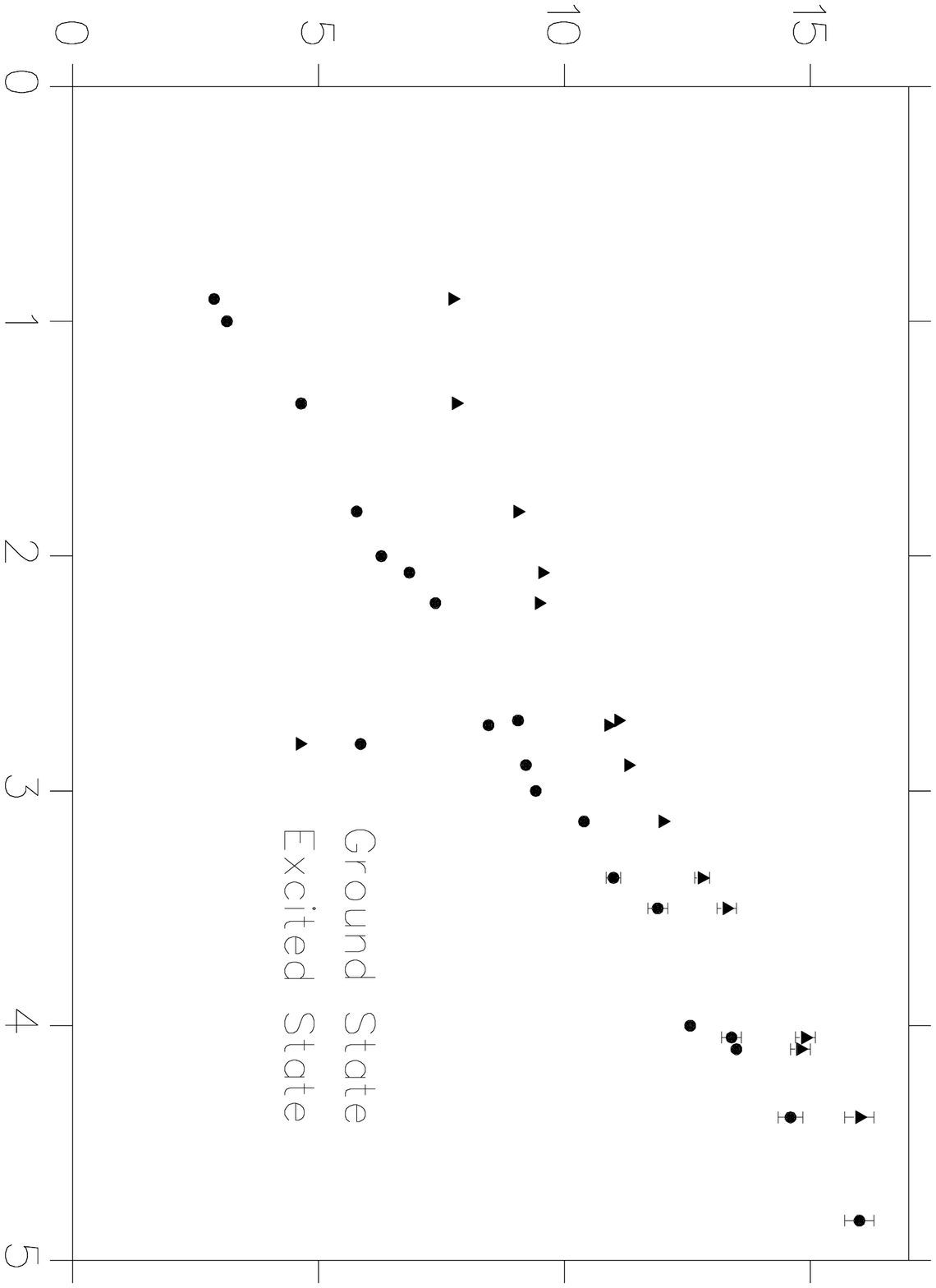}
\end{picture}
\put(-5.5,5){a.)}
\put(-6.5,4.3){2+1 dimensions}
\put(2,5){b.)}
\put(1,4.3){3+1 dimensions}
\begin{picture}(5,6)(-8.5,.8)
\includegraphics{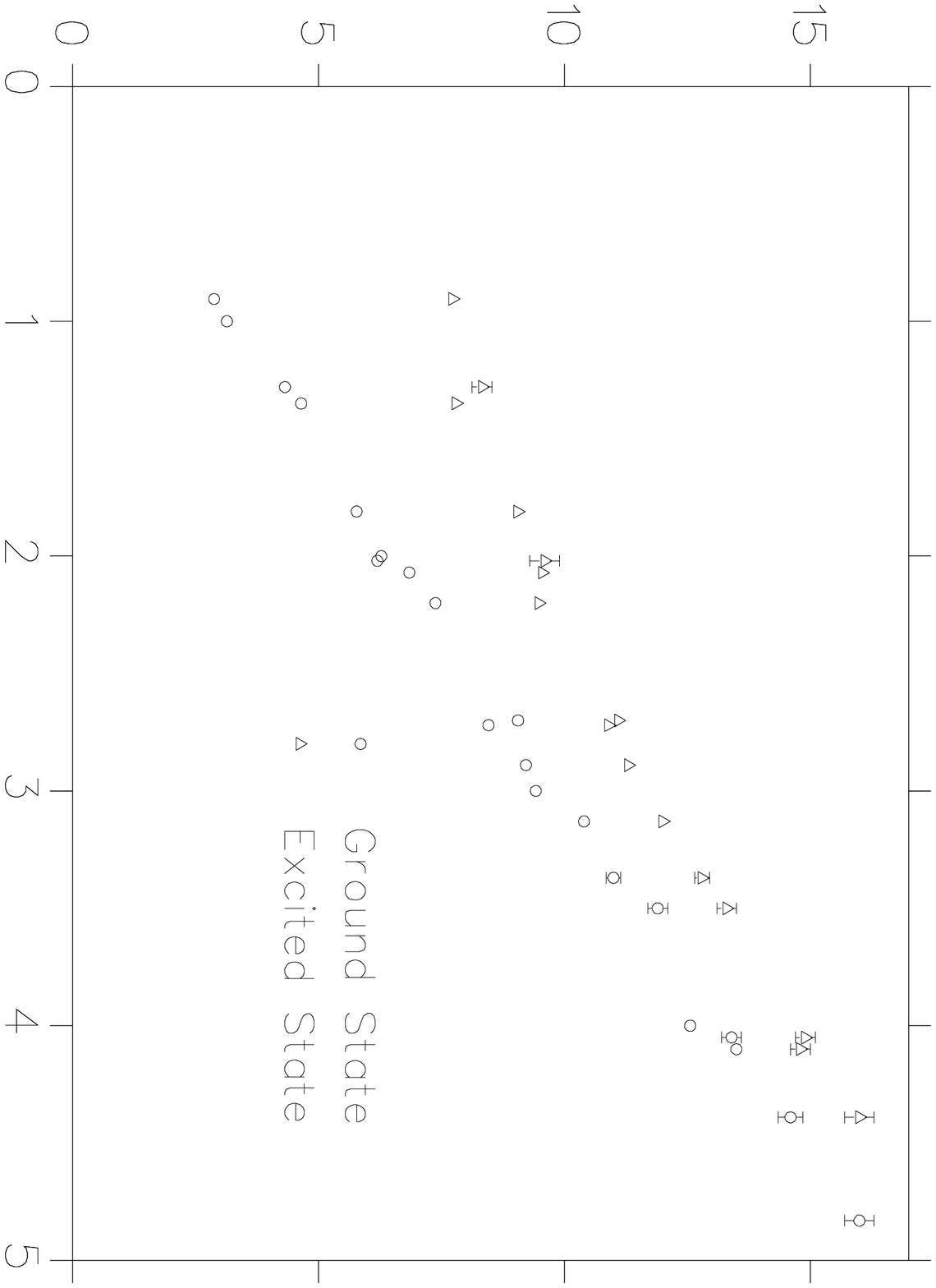}
\end{picture}
\caption{a.) Static $Q\bar{Q}$ potential in the color dielectric formulation of
the $\perp$ lattice in 2+1 dimensions versus the distance 
$r\equiv \protect{\sqrt{x_\perp^2 + x_L^2}}$ between the 
$Q$ and the $\bar{Q}$ . Circles: ground state,
triangles: first adiabatic hybrid state. 
The points represent calculations where the
$Q\bar{Q}$ pair was separated by up to 4 lattice units in
the transverse direction as well as by various separations in the longitudinal 
direction. In 3+1 dimensions, as long as the $\perp$ separation
between the $Q$ and the
$\bar{Q}$ is along one of the $\perp$ axis, the results are identical.
b.) Same as a.), but for 
a $3+1$ dimensional $\perp$ lattice. The transverse separations
between the $Q\bar{Q}$ pair are $(n_x,n_y)=(1,0),\, (2,0),\, (3,0),\,
(1,1)$ and $(2,1),\,$.
}
\label{fig:num}
\end{centering}
\end{figure}
\section{Results in 3+1 Dimensions}
Calculations in 3+1 dimensions introduce two new couplings: a plaquette
coupling and a non-local coupling for states that turn a corner. 
The
ground state potential is found to be approximately
rotationally invariant along the
trajectory $c_p - c_\perp = const.$, where $c_p$ is the plaquette coupling
and $c_\perp$ is the new non-local coupling introduced by going to 3+1
dimensions. The ambiguity along this trajectory is fixed by demanding
that longitudinally and transversely polarized excited (hybrid) states
are degenerate.
Rotational invariance is rather sensitive to the values of the plaquette
coupling and the non-local coupling that turns a corner if they lie off
of the trajectory described above. Roundness is not very sensitive to the
values of the link field mass term, the contact coupling, and the
straight non-local coupling. The optimal set of parameters, in our
units is: $m^2=0.30$, $c_{contact}=-.10$, $c_2 = -.1375$, 
$c_\perp = 0$ and
$c_p=.4875$.

The first excited state of the 3+1 dimensional potential exhibits a
slight anisotropy. This is an expected result of our Fock Space truncation.
However, the shape at intermediate to long distances is encouraging and
it does agree rather well with results obtained on a Euclidean lattice
\cite{lattice}.
\section{Summary and Outlook}
The static $Q\bar{Q}$ potential provides a test of the dynamics
of our model. In all, the rotational invariance of the ground state
potential is striking.
However, the fact that rotational invariance is not
extremely sensitive to all of the parameters in the effective potential
suggests that other observables are needed in order to pin down these
couplings unambiguously \cite{dielectric2}. 

Alternatively, it is conceivable that one could determine
the coefficients of the effective potential on the Euclidean lattice
where one can make the transverse lattice spacing small.
Not only could one be rather confident of the calculations
done from such a theory, but one could also systematically determine
the dependence of observables on high orders in the effective
potential. Such work is currently underway \cite{studs2}.
\end{document}